# Design and Implementation of a GUI based Offline GIFT Tool to exchange data between different systems

Kisor Ray
Techno India Agartala,
An Engineering College affiliated to Tripura University
Maheshkhola, Tripura (West), India

Partha Pratim deb
Techno India Agartala,
An Engineering College affiliated to Tripura University
Maheshkhola, Tripura (West), India

## ABSTRACT
Multiple Choice Questions or MCQs are very important for e-learning. Many MCQ Tools allow us to generate MCQs very easily. However, in most of the cases they are not portable. That means MCQs generated for one system cannot be used for other unless a common format is used. So, collaboration and/or up gradation becomes a time consuming tedious task. In this paper, we will examine how tool could be designed which can produce portable MCQs and that too generating in the laptop and/or desktop without any need for going online.

## 1. INTRODUCTION
MCQs are used very widely across the globe to evaluate students' performance by the various academic institutes and universities starting from the entrance examinations to periodic tutorials, assignments and semester examinations. Corporate houses and public sectors use MCQs for the screening of the prospective employees. These are used for general quiz competitions organized by different entities including the electronic ones. MCQs are widely used by the e-learning portals for quizzing the participants after the end of the courseware to find the level of knowledge being gathered by them. E-learning needs a couple of components, MCQs being one of them, to serve as a knowledge evaluation tool. E-learning offers online courseware and computer assisted assessment thus reducing the burden of managing a large pool of students compared to traditional paper based testing process. Different organizations use different tools (commercial, open source or in-house custom made) to create MCQs. Can we make MCQs produced by one organization portable to another one even though they are not using the similar system? In the countries where regular internet connection is a problem at house hold level due to the poor bandwidth, can we make a tool available to the teachers/faculty members which can enable them to produce MCQs on their laptop/desktop without going to their destination portal using the internet? Can we reduce the need of purchasing more commercial licenses in a situation where the organization is using a commercial product? In the paper we will attempt to bring certain solutions for the problem areas mentioned earlier.

## 2. MOTIVATION
Interest of the students is the key for the success of e-learning courseware. This is one part of the story. The involvement of the teachers/faculty members and ease of use in production of MCQs as well as exchange of the same between different likeminded organizations as an easy portable commodity enhances the chance of bringing up the quality collaboratively. If a tool could be designed and make it available at no cost/free tool which can make this collaboration happen covering the commercial tool to open source that too without any need to use the online portal, may inspire many institute and their respective faculty members to use this tool.

## 3. BACK GROUND
At our organization level, we have Moodle based LMS. Any faculty wants to create MCQs, can use the LMS and generate MCQs as an integral part of the courseware. Now, here is the problem statement: MCQs could be generated but then the Moodle needs to be accessed through the internet! In our country, the bandwidth of the internet is a problem in many places, especially in the northeast. So, accessing the Moodle based LMS is an option but not the best for the obvious reason of connectivity issue. We had tried to evaluate the commercially available Adobe Captivate which generates good looking commercial grade quizzes. But then, we need to have sufficient licenses for all the participating faculty members to run the product in their respective laptops/desktops. Moreover, should we need to exchange the quizzes between our Moodle based LMS and Adobe based Quizzing system, the only option is to go to the Moodle based portal online, generate the quiz and then export it as a GIFT format and/or manually generate (using text editors) quizzes using GIFT formatting (which many will not like specially in absence of IT background since the formatting may look like scripts!) and use this as a feed for the systems which supports this format for example, Moodle and Adobe Captivate. One of the no cost best possible solution is to design, develop and implement a tool which can take Graphical User Interface (GUI) based inputs to create MCQS in the database and capable of converting the inputs as GIFT format. The GIFT formatted output of this tool could be used as input feed to the Moodle and/or Adobe Captivate based system. So, what we see that this new tool:

(1) Can be used by the teachers/faculty members at their desktop/laptop like any GUI driven easy to use tool
(2) No need to hit the target portal which means no need to use the internet while creating the MCQS

| SL | CB1 | CB2 | CB3 | CB4 | Res1_T | Res2_T | Res3_T | Res4_T | |
|---|---|---|---|---|---|---|---|---|---|
| 0 | ☐ | ☐ | ☐ | ☐ |   |   |   |   | F F F F |
| 1 | ☐ | ☐ | ☐ | ☑ | ~ | ~ | ~ | = | F F F T |
| 2 | ☐ | ☐ | ☑ | ☐ | ~ | ~ | = | ~ | F F T F |
| 3 | ☐ | ☐ | ☑ | ☑ | ~ | ~ | = | = | F F T T |
| 4 | ☐ | ☑ | ☐ | ☐ | ~ | = | ~ | ~ | F T F F |
| 5 | ☐ | ☑ | ☐ | ☑ | ~ | = | ~ | = | F T F T |
| 6 | ☐ | ☑ | ☑ | ☐ | ~ | = | = | ~ | F T T F |
| 7 | ☐ | ☑ | ☑ | ☑ | ~ | = | = | = | F T T T |
| 8 | ☑ | ☐ | ☐ | ☐ | = | ~ | ~ | ~ | T F F F |
| 9 | ☑ | ☐ | ☐ | ☑ | = | ~ | ~ | = | T F F T |
| 10 | ☑ | ☐ | ☑ | ☐ | = | ~ | = | ~ | T F T F |
| 11 | ☑ | ☐ | ☑ | ☑ | = | ~ | = | = | T F T T |
| 12 | ☑ | ☑ | ☐ | ☐ | = | = | ~ | ~ | T T F F |
| 13 | ☑ | ☑ | ☐ | ☑ | = | = | ~ | = | T T F T |
| 14 | ☑ | ☑ | ☑ | ☐ | = | = | = | ~ | T T T F |
| 15 | ☑ | ☑ | ☑ | ☑ | = | = | = | = | T T T T |

Fig 1: Master Lookup table for GIFT symbol search and pickup based on the user input

(3) Teachers/faculty members can produce GIFT formatted feed and handover to the system admin
(4) Dissimilar system for example, Moodle and Adobe Captivate could be used with ease by making the data portable
(5) Eliminates the multiple licensing need of commercial software.
(6) All the MCQs created using this tool could be easily stored in a centralized database under different categories, thus making question pools.

There are many different types of MCQs like:
1. True/False
2. Multiple Choice
3. Multiple Response
4. Fill in the Blank
5. Matching
6. Numeric/Number Range
7. Hotspot

Among these 7 different types, we have used GIFT formatting to generate type 1 to type 6. GIFT format allows one to use a text editor to write quiz questions in a simple format that can be imported into Adobe Captivate or Moodle Quiz. This makes it easier to port the quizzes by just producing them using simple text editors like Notepad, Notepad++ under Windows or nano and vi under Linux/Unix operating system. For example the following questions could be generated using simple text editors and following the GIFT formatting.[1][2]

// true-false
::Q1:: 1+1=2 {T}
// multiple choice with specific feedback
::Q2:: What's between orange and green in the spectrum?
{=yellow # correct! ~red # wrong, it's yellow ~blue # wrong, it's yellow}
// fill-in-the-blank
::Q3:: Two plus {=two =2} equals four.
// matching
::Q4:: Which animal eats which food? { =cat -> cat food =dog -> dog food}

// math range question -- note: {#1..5} is the same range
:: Q5:: What is a number from 1 to 5? {#3:2}

There is an online tool (Moodle test creator) [3] which can generate GIFT formatted output, but may not satisfy one of our major objectives of not using the internet at any point of time while creating the MCQs. So, we decided to design ,develop and implement an offline GUI based easy to use tool which can take the quizzes through its normal GUI based inputs and can produce the script like GIFT formatted texts for the purpose of using this with system like Moodle or Adobe Captivate.

## 4. DESIGN CONCEPT

What we have seen in the GIFT formatting example earlier is that the wrong answer corresponds to the symbol '~' and the correct answer is identified by "=" . For the sake of simplicity, if we take questions types as "True/False", "Single Answer", "Multiple Choices" and " "Fill the Blank" , we will find that that the maximum combinations can occur for the types "Multiple Choices" that is 16 such logical combinations can take place which can be the super sets for all other types. Thus, to create our target tool , we need a look up table which stores all the possible combinations and corresponding symbols. Based on the response ( marked as RBs below) by the user (Quiz creator), the master look up table is read (Shown as CBs below) and then the response is then stored in the transaction table along with the actual question and answer text.

```
    Dim rst As DAO.Recordset
    Set db = CurrentDb

    Set rst = db.OpenRecordset (" Select * from Tbl_M_LookUp where CB1 = " & Me.RB1 & "AND CB2 = " & Me.RB2 & "AND CB3 = " & Me.RB3 & " AND CB4 = " & Me.RB4 & " ;", dbOpenDynaset, dbSeeChanges)

        If Not rst.EOF Then
         MsgBox "Debug - Got them here at C3" & rst.RecordCount
         Me.TxTR1_T = rst ("Res1_T")
         Me.TxTR2_T = rst ("Res2_T")
         Me.TxTR3_T = rst ("Res3_T")
         Me.TxTR4_T = rst ("Res4_T")
```

```
        Else

           MsgBox "Problem in Lookup! "

        End If

        rst. Close
        db. Close
```

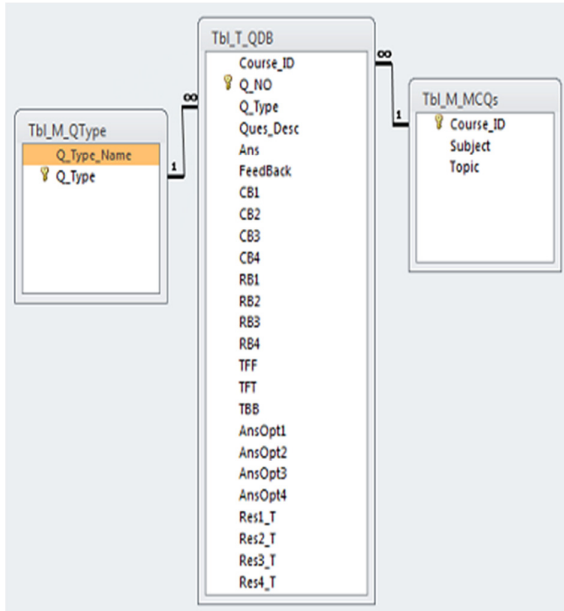

Fig 2: Question Database Table (Transaction) and relationship with the master tables

In the database design, three master tables and one transaction table is used to retrieve and store data.

Table 1: Details of the Database Tables

| SL No. | Table Name | Table Type | Purpose |
|---|---|---|---|
| 1 | Tbl_M_Q_Type | Master | Stores Question types (Eg. Multiple Choice, True/False etc.) |
| 2 | Tbl_M_MCQs | Master | Stores Course ID and Subjects |
| 3 | Tbl_M_Lookup | Master | Stores GIFT Symbols to Map with the user responses |
| 4 | Tbl_T_QDB | Transaction | Stores Quiz questions from the GUI based front end |

## 5. HOW IT WORKS
The course admin creates the Course ID and subject name which is inserted into the master table Tbl_M_MCQs. The end user uses the Course ID and inputs test questions using GUI based easy to use front end. After questions are entered into the Question database, they could be easily exported.

The export will produce a GIFT formatted output. For example:
```
        Course ID    C300
        Ques type    3
         :1:
            Water is a compound of two different elements. They are:
            {
              ~    Nitrogen
              =    Oxygen
              ~    Carbon Di-Oxide
              =    Hydrogen
              #    Oxygen and Hydrogen
            }
```

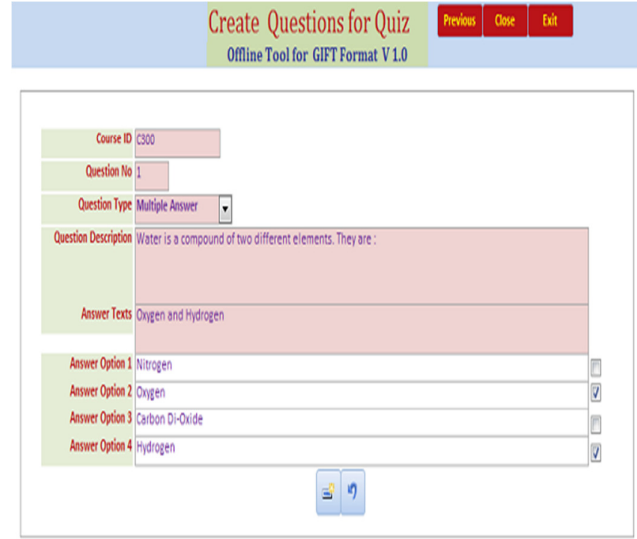

Fig 3: GUI based front end to take input from the faculty members/teachers

When this GIFT formatted question is imported into Adobe Captivate, we can get the screen below:

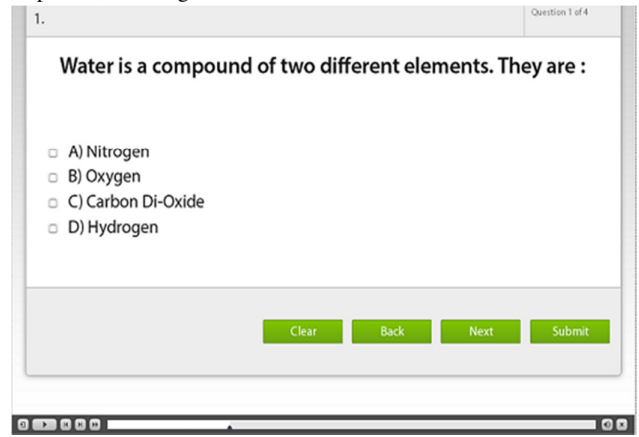

Fig 4 : Imported Gift formatted Quiz creates question for the end users (students) using Adobe Captivate

## 6. IMPLEMENTATION
We have used MS Access to create the GUI front end and PostGreSQL as the backend database to store the data in the backend keeping in mind to create a large repository of question bank categorized under different Course ID and subjects. Any technology could be used to create the GUI using the database schema and symbol lookup table suggested through our project. We have also developed code to handle large objects (lo) using PostGreySQL ( beyond the scope of this paper) with an objective to use interactive images while creating the quizzes.

## 7. CONCLUSION

Our desktop tool is a full-fledged easy to use GUI based tool which does not need to use the internet while creating the quizzes and the output could be easily used (by importing into) with dissimilar systems like Adobe Captivate and Moodles to serve the end users like students. We genuinely felt that we need a good desktop based easy to use tool owing to the lack of good internet bandwidth. We failed to find any such tool which can serve us under our situation. So, we finally decided to design, develop and implement this custom tool per our need. "GIFT allows someone to use a text editor to write multiple-choice, true-false, short answer, matching, missing word and numerical questions in a simple format that can be imported to a computer-based quizzes. The content is an UTF-8-encoded text file".[4], we wanted to provide the users more comfort and control by allowing them to create GIFT using a standalone GUI tool.

.